\documentclass[12pt]{iopart}

\usepackage{epsfig,graphics}
\begin{document}
\title {First-forbidden transitions and stellar $\beta$-decay
rates of Zn and Ge isotopes}
\author{Jameel-Un Nabi$^1$ \footnote{Corresponding author}, Necla Cakmak$^2$, Sabin Stoica$^3$ and Zafar Iftikhar$^1$ }
\address{1 Faculty of Engineering Sciences, GIK Institute of Engineering Sciences and
Technology, Topi 23640, Khyber Pakhtunkhwa, Pakistan
\\ 2  Department of Physics, Karabuk University, Karabuk, Turkey
\\ 3 Horia Hulubei Foundation, P. O. Box MG-12, 071225, Magurele,
Romania} \eads{jameel@giki.edu.pk}

\begin{abstract}
First-forbidden (FF) charge-changing transitions become relatively
important for nuclei as their proton number increases. This is
because the strength of allowed Gamow-Teller (GT) transitions
decreases with increasing Z. The FF transitions play an important
role in reducing the half-lives as against those calculated from
taking the GT transitions alone into account. In this paper we
calculate allowed GT as well as $0^{+} \rightarrow 0^{-}$ and $0^{+}
\rightarrow 2^{-}$ transitions for neutron-rich Zn and Ge isotopes.
Two different pn-QRPA models were used with a schematic separable
interaction to calculate GT and FF transitions. Half-lives
calculated after inclusion of FF transitions were in excellent
agreement with the experimental data. Our calculations were also
compared to previous QRPA calculations and were found to be in
better agreement with measured data. Stellar $\beta$-decay rates
were calculated for these nuclei including allowed GT and unique FF
transitions for astrophysical applications. $^{86,88}$Ge has a
sizeable contribution to the total stellar rate from unique FF
transitions.
\end{abstract}
 \maketitle


\section{Introduction}
The subject of $\beta$-decay of neutron-rich nuclei and their half
lives have gained interest in the past few years. This is partly due
to incoming results from the new generation of radioactive ion-beam
facilities and partly due to their vital role in solving problems
related to the stellar evolution.  The eventual abundance of any
stable nucleus depends strongly on the $\beta$-decay half life of
its neutron rich forebear. Electron-neutrino captures could not only
amplify the effect of $\beta$-decays in neutron rich environments
but the subsequent neutrino induced neutron spallation can also
contribute towards changing the $r$-abundance distribution pattern
\cite{McL97}. Stellar weak interaction processes, for densities
$\rho \le$ 10$^{11}$ g/cm$^{3}$, are dominated by Gamow-Teller (GT)
and, if applicable, also by Fermi transitions. Forbidden transition
contributes generously, as one moves away from the line of
stability, when the electron chemical potential reaches a value of
 $\sim$~30 MeV.  In case of neutron-rich nuclei, the
first-forbidden (FF) $\beta$-decay rates may become important due to
the enlarged phase space for these transitions for densities  $\rho
\leq 10^{11} g/cm^{3}$ \cite{Nab13}. Forbidden transitions
contribute significantly to the total half-life for nuclei crossing
the closed N and Z shells, specially for N $<$ 50 in $^{78}$Ni
region. At the same time the FF transitions become relatively
important for nuclei with large Z. The contribution from the allowed
GT transitions gets smaller for such nuclei and bears consequences
for the nucleosynthesis calculations \cite{Suz11}.

Primarily due to the lack of experimental data, majority of the
$\beta$-decay rates for neutron rich nuclei have been investigated
using theoretical models. The proton neutron quasi-particle random
phase approximation (pn-QRPA) model has been widely used in studies
of nuclear $\beta$-decay properties. In this microscopic model,
construction of a quasi-particle basis is first performed with a
pairing interaction, and then the RPA equation is solved with GT
residual interaction. The pn-QRPA model was developed by Halbleib
and Sorensen \cite{Hal67} by generalizing the usual RPA to describe
charge-changing transitions. The hybrid version of RPA model was
developed by M\"{o}ller and his coworkers \cite{Mol03}. They
combined the pn-QRPA model with the Gross Theory of the FF decay.
For studying deformed nuclei and nuclei with odd nucleons many
changes were made to the pn-QRPA model
\cite{Ran73,Mol90,Ben88,Mut89,Mut92}. It was Nabi and Klapdor who
used the pn-QRPA model, for the first time, to calculate stellar
weak rates \cite{Nab99a,Nab99,Nab04}. A methodical study of the
total $\beta$-decay half-lives and delayed neutron emission
probabilities using the pn-QRPA model, taking into account the GT
and FF transitions , can be further examined in \cite{Hom96}.

In this paper an attempt is made to calculate the FF $\beta$-decay
rates for neutron-rich even-even zinc ($^{76-82}$Zn) and germanium
isotopes ($^{84-88}$Ge) using the pn-QRPA model. Motivation of the
current work came partly from the work of \cite{Nab13} where the
authors expressed their interest to include rank 0 contribution to
first forbidden decay rates for a better agreement with the measured
data. In \cite{Nab13} the  unique first-forbidden (U1F)
$\beta$-decay rates (rank 2) and total $\beta$-decay half-lives for
$^{72-78}$Ni were calculated for the first time using the pn-QRPA
model in stellar environment. The aim of this paper is to highlight
the importance of FF transitions in the calculation of $\beta$-decay
rates of neutron-rich Zn and Ge isotopes. The formalism of allowed
GT, FF and U1F transition rates is given in Section~2 of this paper.
Section~3 show our calculation of allowed and FF charge-changing
transitions for Zn and Ge isotopes. Here we compare our calculated
half-lives with experimental and other theoretical calculations. We
also present our calculation of stellar $\beta$-decay rates in this
section. We finally conclude our findings in Section~4.

\section{Formalism}
The theory of allowed and FF $\beta$-decay transitions
 is well-established \cite{Gre08,Sch66,WuC66,Com73}. The allowed
$\beta$-decay is simple to calculate but the FF decay shows a far
wider spectrum both in lepton kinematics and in nuclear matrix
elements. All our calculations for allowed and FF $\beta$-decay
rates were performed within the framework of the pn-QRPA model. Two
different pn-QRPA models were used to calculate allowed and
forbidden $\beta$-decay rates. The first pn-QRPA model considered
only spherical nuclei using the Woods-Saxon potential basis and is
referred to as pn-QRPA(WS) in this paper. The transition
probabilities in this model were calculated within the $\xi$
approximation ($\xi$ is a dimensionless parameter representing the
magnitude of the Coulomb energy and is approximated by $1.2 Z
A^{-1/3}$). Calculation of rank 0 FF transitions
(0$^{+}$$\rightarrow$0$^{-}$) and U1F transitions
(0$^{+}$$\rightarrow$2$^{-}$) was done within the pnQRPA(WS)
formalism. Details of this model can be seen from \cite{Cak10}. For
the same model, allowed GT transitions were calculated using the
Pyatov method (PM) \cite{Pya77} to solve the RPA equation. The
second pn-QRPA model employed a deformed Nilsson basis and is
referred to as pn-QRPA(N) in this paper. A separable interaction was
used both in particle-particle and particle-hole channels which
transformed the eigenvalue equation to an algebraic equation of
fourth order (for further details of solution see \cite{Mut92}).
Deformation of nuclei was taken into account in the pn-QRPA(N)
model. Allowed GT and U1F transitions were calculated within the
pn-QRPA(N) formalism. The theoretical models used in our calculation
make it stand apart from previous calculations (e.g.
\cite{Bor05,Mol97}). The continuum-QRPA framework employed by
\cite{Bor05} included a self-consistent mean-field potential (for
ground state) and a universal effective $NN$ interaction (for
excited states) for calculation of GT and FF transitions. Both
potentials originated from the unique nuclear energy-density
functional. On the other hand \cite{Mol97} used a folded-Yukawa
single-particle Hamiltonian for calculation of $\beta$-decay rates.
They incorporated pairing, deformed Nilsson basis and a separable
interaction akin to our calculation. However they did not consider
particle-particle interaction. Nor did they calculate FF decays as
done in current work.

The axial-vector coupling constant $g_{A}$ is renormalized (mainly
due to truncations in the nuclear structure calculations but also
from the interference of non-nucleonic degrees of freedom). Ref.
\cite{Suh13} suggested that $g_{A} \le  0.80$ in the pn-QRPA
calculation brings the theoretical and experimental decay value in
close correspondence for $A$ = 100, 116 and 128 isobaric chains. It
was also commented in this paper that a similar type of quenching is
obtained in the interacting shell model and interacting boson model.
(See also \cite{Suh14} for accessing $g_{A}$ value close to $0.6 \pm
0.2$ in the pn-QRPA calculation). For U1F transitions authors in
\cite{Eji14} employed a much tighter quenching factor of 0.5. In
order to compare our results with experimental data and previous
calculations, we introduced a quenching factor of $0.6$
\cite{Vet89,Cak14,Cak14a} in both pn-QRPA(WS) and pn-QRPA(N) models
as  also used in shell model calculations. In future we would like
to explore the effect of using different quenching factor for
allowed, FF and U1F transitions in pn-QRPA(WS) and pn-QRPA(N)
models. The same quenching factor was employed later to calculate
stellar $\beta$-decay rates.

Below we describe briefly  the formalism used to calculate GT, FF
and U1F transitions using the pnQRPA(WS) model. We later present the
formalism for GT and U1F $\beta$-decay rates using the pnQRPA(N)
model.

\subsection{Allowed GT, FF and U1F transitions using the pn-QRPA(WS) model}
The model Hamiltonian which generates the spin-isospin dependent
vibration modes with $I^{\pi}=0^{-}, 2^{-}$ in odd-odd nuclei in
quasi boson approximation is given by
\begin{eqnarray}
\hat{H}=\hat{H}_{sqp}+\hat{h}_{ph}+\hat{h}_{pp},
\end{eqnarray}
where the single quasi-particle (sqp) Hamiltonian of the system is
given by
\begin{eqnarray}
\hat{H}_{sqp}=\sum_{j_{k}}\varepsilon_{j_{k}}\alpha^{\dag}_{j_{\tau}m_{\tau}}\alpha_{j_{\tau}m_{\tau}}
(\tau=p,n).
\end{eqnarray}
In above equation $\varepsilon_{j_{k}}$ is single quasi-particle
energy of the nucleons with angular momentum ${j_{k}}$, and
$\alpha^{\dag}_{j_{\tau}m_{\tau}}$ and ($\alpha_{j_{\tau}m_{\tau}}$)
is the quasi-particle creation (annihilation) operators.

The $\hat{h}_{ph}$ and $\hat{h}_{pp}$ are the spin-isospin effective
interaction Hamiltonians, which generates $0^{-}$ and $2^{-}$
vibration modes in particle-hole and particle-particle channel,
respectively. To save space we are not reproducing the detailed
formalism. Interested readers are referred to \cite{Cak10}.

The transitions probabilities $B^{FF}(0^{+}\longrightarrow0^{-}_{i},
\beta^{-})$ in $\xi$ approximation are given by \cite{Boh69}
\begin{eqnarray}
B^{FF}(0^{+}\longrightarrow0^{-}_{i},\beta^{-})=|<0^{-}_{i}\|M^{FF}_{\beta^{-}}\|0^{+}>^{2},
\end{eqnarray}
where
\begin{eqnarray}
M^{FF}_{\beta^{-}}= - M^{-}(\rho_{A},\lambda=0)\nonumber
\end{eqnarray}
\begin{eqnarray}
-i\frac{m_{e}c}{\hbar}\xi M^{-}(j_{A},\kappa=1,\lambda=0),
\end{eqnarray}

\begin{eqnarray}
M^{-}(\rho_{A},\lambda=0)=\frac{g_{A}}{c\sqrt{4\pi}}\Sigma_{k}t_{-}(k)(\vec{\sigma}(k)\cdot\vec{V}(k)).
\end{eqnarray}
\begin{eqnarray}
M^{-}(j_{A},\kappa=1,\lambda=0)=g_{A}\Sigma_{k}t_{-}(k)r_{k}(Y_{1}(r_{k})\sigma_{k})_{0}.
\end{eqnarray}

The transitions probabilities $B(0^{+}\longrightarrow2^{-}_{i},
\beta^{-})$ in $\xi$ approximation are given by \cite{Boh69}
\begin{eqnarray}
B^{U1F}(0^{+}\longrightarrow2^{-}_{i},\beta^{-})=|<2^{-}_{i}\|M^{U1F}_{\beta^{-}}\|0^{+}>^{2},
\end{eqnarray}
where
\begin{eqnarray}
M^{U1F}_{\beta^{-}}= M^{-}(j_{A},\kappa=1,\lambda=2,\mu),\nonumber
\end{eqnarray}
and the $M^{-}(j_{A},\kappa=1,\lambda=2,\mu)$ are non-relativistic
first forbidden $\beta$ decay multipole operators \cite{Boh69},
\begin{eqnarray}
M^{-}(j_{A},\kappa=1,\lambda=2,\mu)=g_{A}\Sigma_{k}t_{-}(k)r_{k}{Y_{1}(r_{k})\sigma(k)}_{2\mu}.
\end{eqnarray}
All symbols have there usual meanings.

The ft values are given by the following expression:
\begin{eqnarray}
(ft)_{\beta^{-}}=\frac{D}{(g_{A}/g_{V})^{2}4\pi
B^{FF(U1F)}(I_{i}\longrightarrow I_{f}, \beta^{-})},
\end{eqnarray}
where
\begin{eqnarray}
D=\frac{2\pi^{3}\hbar^{2}ln2}{g_{V}^{2}m_{e}^{5}c^{4}}=6250sec,~~~\frac{g_{A}}{g_{V}}=-1.24.\nonumber
\end{eqnarray}

In order to calculate the allowed GT transitions in the pn-QRPA(WS)
model we note that the GT operators are given by
\begin{eqnarray}
G_{\mu}^{+}=\sum_{i=1}^{A}\sigma_{\mu}(i)t_{+}(i),~~~G_{\mu}^{-}=(-1)^{\mu}\sum_{i=1}^{A}\sigma_{-\mu}(i)t_{-}(i),
\nonumber
\end{eqnarray}
\begin{eqnarray}
~~G_{\mu}^{-}=(G_{\mu}^{+})^{\dag}.
\end{eqnarray}
The commutation condition between the total Hamiltonian and GT
operator can be described as
\begin{eqnarray}
[H,G_{\mu}^{\pm}]=[V_{1}+V_{c}+V_{\vec{l}\vec{s}},G_{\mu}^{\pm}],
\end{eqnarray}
where $V_{1}$, $V_{c}$, and $V_{\vec{l}\vec{s}}$ are isovector,
Coulomb, and spin-orbit interaction potentials, respectively. We
considered a system of nucleons in a spherical symmetric average
field with pairing forces. The corresponding quasi-particle
Hamiltonian is given by Eq.~2 and does not commute with the GT
operators. The broken symmetry is then restored using an effective
interaction employing the Pyatov method. Details of solution of
allowed GT formalism can be seen in \cite{Cak12}.

The ft values for the allowed GT $\beta$ transitions were finally
calculated using
\begin{eqnarray}
ft=\frac{D}{(\frac{g_{A}}{g_{V}})^{2}4\pi B^{GT}(I_{i}\rightarrow
I_{f},\beta^{-})},
\end{eqnarray}
where the reduced matrix elements of GT transitions are given by
\begin{eqnarray}
B^{GT}(I_{i}\rightarrow
I_{f},\beta^{-})=\sum_{\mu}|\langle1_{m}^{+},\mu|G_{\mu}^{-}|0_{g.s.}^{+}\rangle|^{2}.
\end{eqnarray}

\subsection{Allowed GT, U1F transitions and stellar $\beta$-decay rates using the pn-QRPA(N) model}
In the pn-QRPA(N) formalism \cite{Mut92}, proton-neutron residual
interactions occur as particle-hole (characterized by interaction
constant $\chi$) and particle-particle (characterized by interaction
constant $\kappa$) interactions. We used a schematic separable
interaction (as in the case of pn-QRPA(WS) model). Details of the
separable potential may be seen from \cite{Nab13} (and references
therein). The advantage of using these separable GT forces is that
the QRPA matrix equation reduces to an algebraic equation of fourth
order, which is much easier to solve as compared to full
diagonalization of the non-Hermitian matrix of large dimensionality
\cite{Mut92,Hom96}.

Essentially we first constructed a quasiparticle basis (defined by a
Bogoliubov transformation) with a pairing interaction, and then
solved the RPA equation with a schematic separable GT residual
interaction. The single particle energies were calculated using a
deformed Nilsson oscillator potential with a quadratic deformation.
The pairing correlation was taken into account in the BCS
approximation using constant pairing forces. The BCS calculation was
performed in the deformed Nilsson basis for neutrons and protons
separately. The formalism for calculation of allowed $\beta$-decay
rates in stellar matter using the pn-QRPA(N) model can be seen in
detail from Refs. \cite{Nab99,Nab04}. Below we describe briefly the
necessary formalism to calculate the  U1F $\beta$-decay rates.

The nuclear matrix elements of the separable forces which appear in
RPA equation are given by

\begin{equation}
V^{ph}_{pn,p^{\prime}n^{\prime}} = +2\chi
f_{pn}(\mu)f_{p^{\prime}n^{\prime}}(\mu),
\end{equation}

\begin{equation}
V^{pp}_{pn,p^{\prime}n^{\prime}} = -2\kappa
f_{pn}(\mu)f_{p^{\prime}n^{\prime}}(\mu),
\end{equation}

where
\begin{equation}
f_{pn}(\mu)=<j_{p}m_{p}|t_{-}r[\sigma Y_{1}]_{2\mu}|j_{n}m_{n}>,
\end{equation}
is a single-particle U1F transition amplitude (the symbols have
their normal meaning). Note that $\mu$ takes the values
$\mu=0,\pm1$, and $\pm2$ (for allowed decay rates $\mu$ only takes
the values $0$ and $\pm1$), and the proton and neutron states have
opposite parities \cite{Hom96}. Interaction constant $\chi$ was
taken to be $4.2/A$ MeV for allowed and $56.16/A$ MeV fm$^{-2}$ for
U1F transitions, assuming a $1/A$ dependence \cite{Hom96}. The other
interaction constant $\kappa$ was taken to be zero. These values of
$\chi$ and $\kappa$ best reproduced the measured half-lives.

Deformation of the nuclei was calculated using
\begin{equation}
\delta = \frac{125(Q_{2})}{1.44 (Z) (A)^{2/3}},
\end{equation}
where $Z$ and $A$ are the atomic and mass numbers, respectively and
$Q_{2}$ is the electric quadrupole moment taken from \cite{Mol81}.
Q-values were taken from the mass compilation of Audi et al.
\cite{Aud12}.

We are currently working on the calculation of rank 0 FF transition
phase space factors at finite temperatures. This would be treated as
a future assignment and currently we are  able to calculate only the
phase space factors for rank 2 U1F transitions under stellar
conditions. The U1F stellar $\beta$-decay rates from the
$\mathit{i}$th state of the parent to the $\mathit{j}$th state of
the daughter nucleus is given by

\begin{equation}
\lambda_{ij}^{\beta} =
\frac{m_{e}^{5}c^{4}}{2\pi^{3}\hbar^{7}}\sum_{\Delta
J^{\pi}}g^{2}f(\Delta J^{\pi};ij)B(\Delta J^{\pi};ij),
\end{equation}
where $f(\Delta J^{\pi};ij)$ and $ B(\Delta J^{\pi};ij)$ are the
integrated Fermi function and the reduced transition probability for
$\beta$-decay, respectively, for the transition $i \rightarrow j$
which induces a spin-parity change $\Delta J^{\pi}$. In Eq.~18  $g$
is the weak coupling constant which takes the value $g_{V}$ or
$g_{A}$ according to whether the $\Delta J^{\pi}$ transition is
associated with the vector or axial-vector weak-interaction. The
phase-space factors $f(\Delta J^{\pi};ij)$ are given as integrals
over the lepton distribution functions and hence are sensitive
functions of the temperature and density in stellar interior. The
$B(\Delta J^{\pi};ij)$ are related to the U1F weak interaction
matrix elements stated earlier.

For the U1F transitions the integral can be obtained as

\begin{eqnarray}
f = \int_{1}^{w_{m}} w \sqrt{w^{2}-1}
(w_{m}-w)^{2}[(w_{m}-w)^{2}F_{1}(Z,w) \nonumber\\
+ (w^{2}-1)F_{2}(Z,w)] (1-G_{-}) dw,
\end{eqnarray}
where $w$ is the total kinetic energy of the electron including its
rest mass and $w_{m}$ is the total $\beta$-decay energy ($ w_{m} =
m_{p}-m_{d}+E_{i}-E_{j}$, where $m_{p}$ and $E_{i}$ are mass and
excitation energies of the parent nucleus, and $m_{d}$ and $E_{j}$
of the daughter nucleus, respectively). $G_{-}$ are the electron
distribution functions.

The Fermi functions, $F_{1}(\pm Z,w)$ and $F_{2}(\pm Z,w)$ appearing
in Eq.~(19) were calculated according to the procedure adopted by
\cite{Gov71}.

There is a finite probability of occupation of parent excited states
in the stellar environment as a result of the high temperature in
the interior of massive stars. Weak decay rates then also have a
finite contribution from these excited states. The occupation
probability of a state $i$ is calculated on the assumption of
thermal equilibrium,

\begin{equation} P_{i} = \frac {exp(-E_{i}/kT)}{\sum_{i=1}exp(-E_{i}/kT)}, \end{equation}
where $E_{i}$ is the excitation energy of the  $i^{th}$ parent
state. The rate per unit time per nucleus for stellar $\beta$-decay
process is finally given by
\begin{equation} \lambda^{\beta} = \sum_{ij}P_{i}
\lambda_{ij}^{\beta}.
\end{equation}
The summation over all initial and final states were carried out
until satisfactory convergence in the rate calculations was
achieved. We note that due to the availability of a huge model space
(up to 7 major oscillator shells) convergence was easily achieved in
our rate calculations for excitation energies well in excess of 10
MeV (for both parent and daughter states).

\section{Results and Comparison}
\begin{figure}[htbp]
\begin{center}
\includegraphics[width=1.0\textwidth]{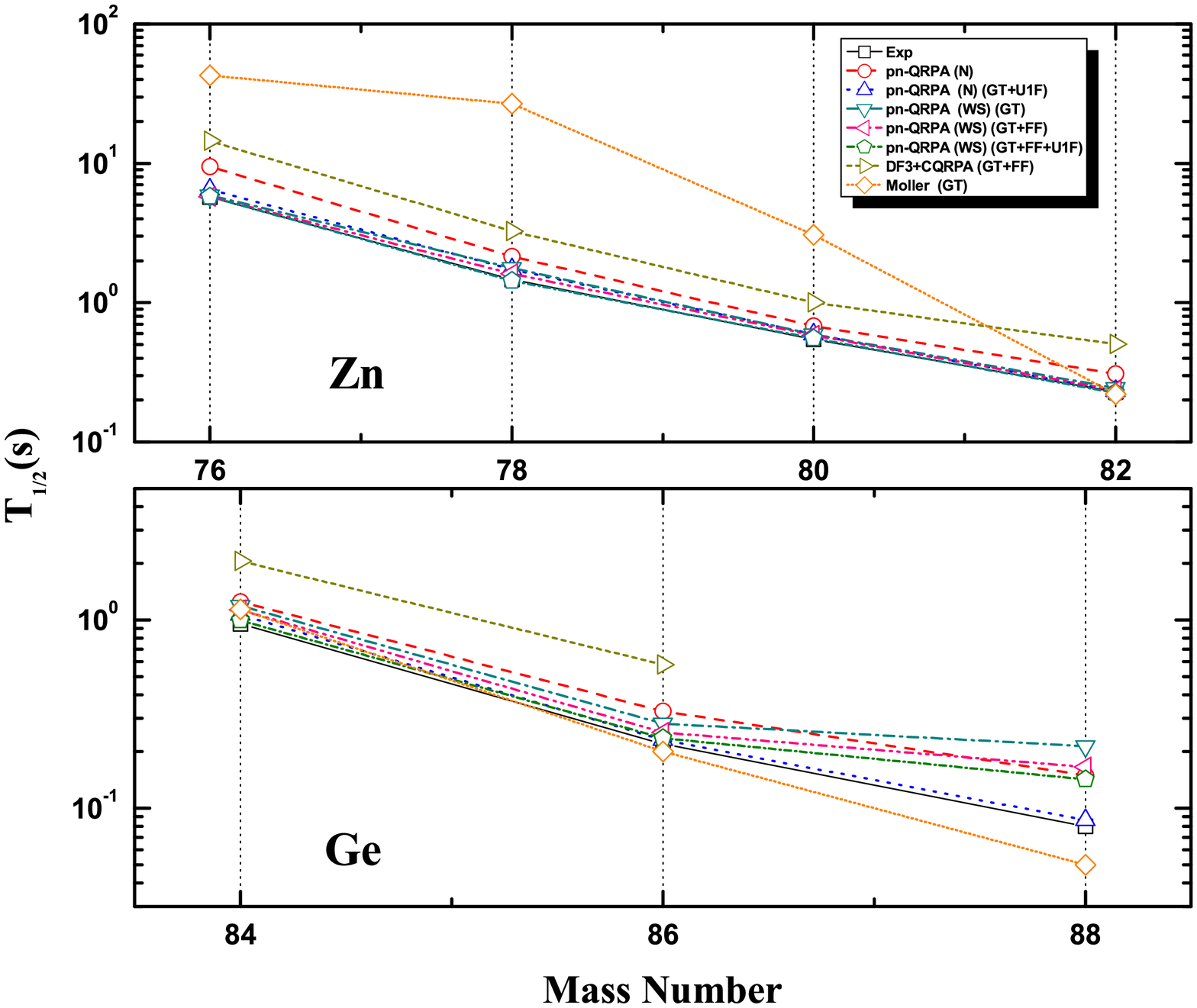}
\caption{\scriptsize Total $\beta$-decay half-lives for Zn and Ge
isotopes calculated from the pn-QRPA(N) and pn-QRPA(WS) models (this
work) including only the allowed (GT), allowed  plus unique
first-forbidden (GT+U1F) and allowed plus first-forbidden plus
unique first-forbidden (GT+FF+U1F) transitions, in comparison with
experimental data \cite{Aud12}, the DF3+CQRPA \cite{Bor05}
calculation and QRPA calculation by \cite{Mol97}.}\label{figure1}
\end{center}
\end{figure}

In order to check the improvement in pn-QRPA calculations brought by
incorporation of FF contribution, we compare our calculated
$\beta$-decay half-lives with experimental data and other model
calculations in Fig.~\ref{figure1}. The upper panel shows result for
Zn isotopes whereas the lower panel for Ge isotopes. All
experimental half-lives were taken from the recent atomic mass
evaluation data \cite{Aud12}. We show results of our calculated
allowed GT calculations alone and those including the FF
contribution. Shown also is the self-consistent density functional +
continuum quasiparticle random phase approximation (DF3 + CQRPA)
calculation of \cite{Bor05} including the FF contribution. The QRPA
calculation by M\"{o}ller and collaborators \cite{Mol97} including
deformation of nucleus and folded-Yukawa single-particle potential
is also presented in Fig.~\ref{figure1}. It is to be noted that
\cite{Mol97} did not calculate FF contribution. It is noted from
Fig.~\ref{figure1} that FF contribution brings substantial
improvement in our pn-QRPA calculated half-lives. The pn-QRPA(N)
calculation including deformation has a sizeable contribution from
FF decay. The pn-QRPA(WS) results are in better agreement with
measured data for Zn isotopes (upper panel) whereas the pn-QRPA(N)
shows overall best agreement with experimental data. For Zn and Ge
isotopes, the DF3+CQRPA results are roughly a factor two bigger than
measured data. GT calculation of \cite{Mol97} are even bigger for
obvious reason. However the agreement is excellent for the case of
$^{82}$Zn. Similarly in lower panel the agreement is outstanding for
the case of  $^{88}$Ge. It may be concluded that the QRPA
calculation of M\"{o}ller et al. gets in better agreement with
measured data as the nucleus becomes more neutron-rich. DF3+CQRPA
did not perform calculation for $^{88}$Ge. It is noted that pn-QRPA
model calculations (with FF contribution) are in excellent agreement
with the experimentally determined half-lives of Zn and Ge isotopes.
The pn-QRPA calculations are expected to give reliable half-lives
for nuclei close to neutron-drip line for which no experimental data
is available.

\begin{figure}[htbp]
\begin{center}
\includegraphics[width=1.0\textwidth]{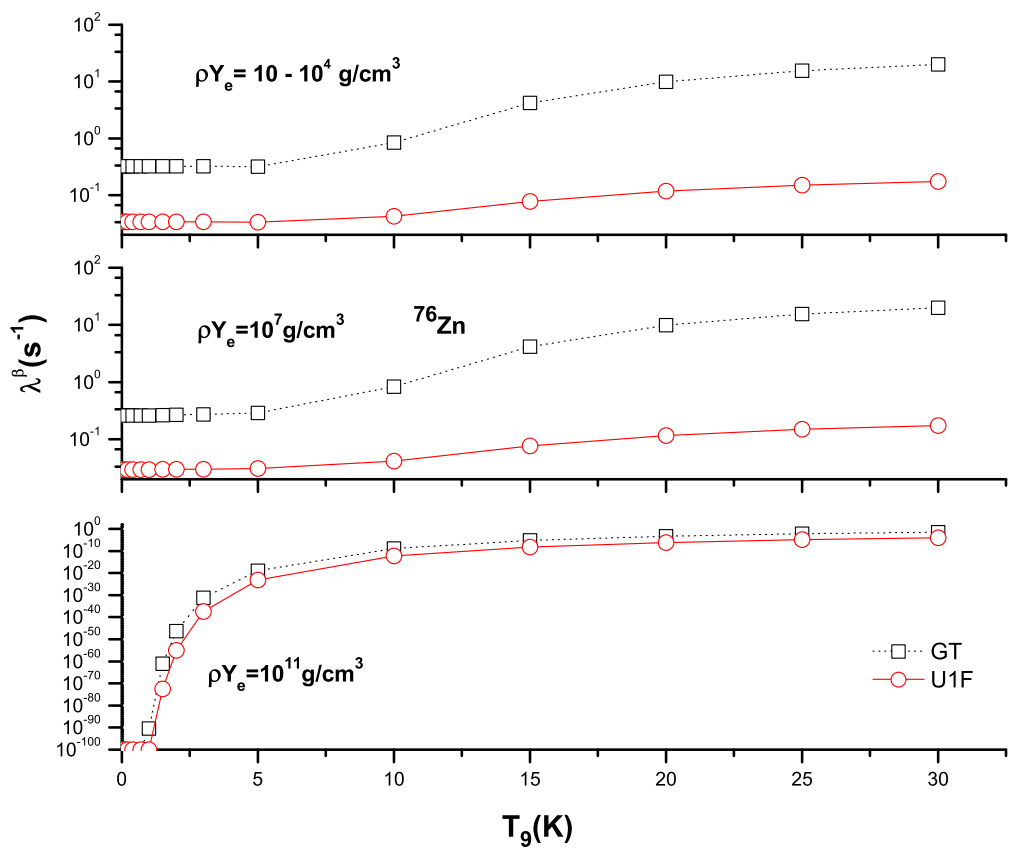}
\caption{\scriptsize Allowed (GT) and  unique first-forbidden (U1F)
$\beta$-decay rates of $^{76}$Zn as a function of temperature for
different selected densities. All $\beta$ decay rates are given in
units of sec$^{-1}$. Temperatures (T$_{9}$) are given in units of
10$^{9}$ K.}\label{figure2}
\end{center}
\end{figure}

\begin{figure}[htbp]
\begin{center}
\includegraphics[width=1.0\textwidth]{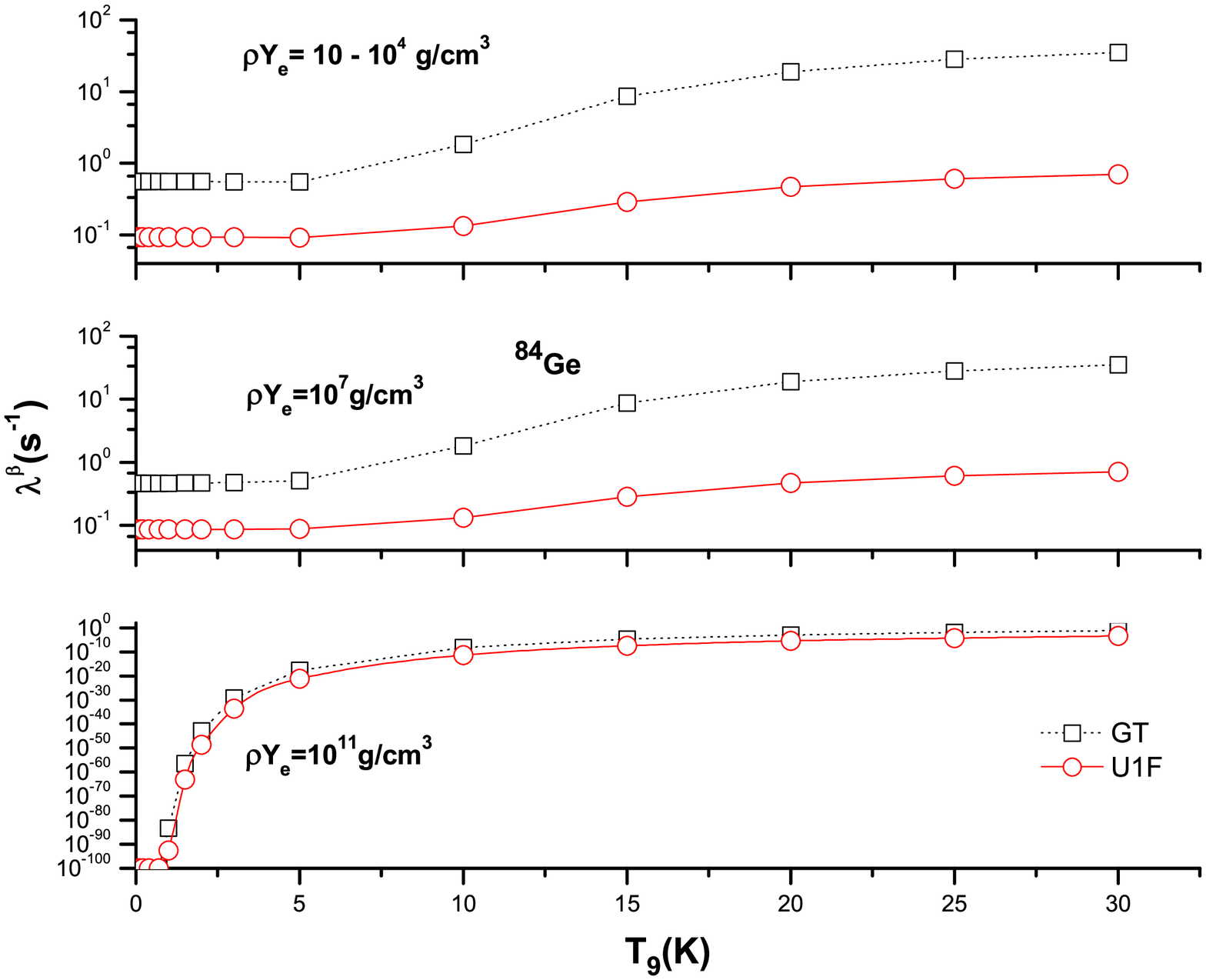}
\caption{\scriptsize Same as Fig.~\ref{figure2} but for
$^{84}$Ge.}\label{figure3}
\end{center}
\end{figure}

\begin{figure}[htbp]
\begin{center}
\includegraphics[width=1.0\textwidth]{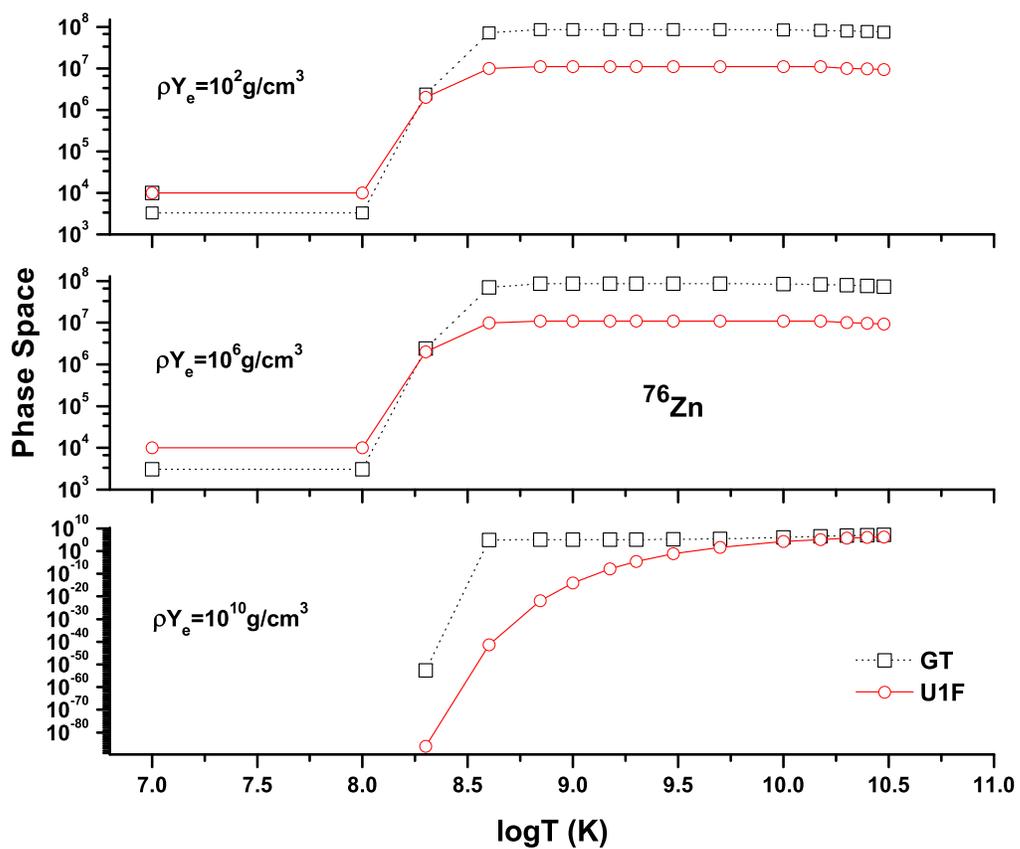}
\caption{\scriptsize Comparison of calculated phase spaces for
allowed and U1F transitions for $^{76}$Zn as a function of stellar
temperatures and densities.} \label{figure4}
\end{center}
\end{figure}

After achieving excellent  comparison with the experimental
half-lives, we next proceeded to calculate allowed GT and forbidden
$\beta$-decay rates in stellar environment.  Allowed and U1F
$\beta$-decay rates were calculated for densities in the range
10~-~10$^{11}$g/cm$^{3}$ and temperature range 0.01 $\leq$ T$_{9}$
$\leq$ 30 (T$_{9}$ gives the stellar temperature in units of
10$^{9}$ K) for Zn and Ge isotopes. Figs.~\ref{figure2}
and~\ref{figure3} show three panels depicting pn-QRPA(N) calculated
allowed and U1F $\beta$-decay rates for temperature range 0.01
$\leq$ T$_{9}$ $\leq$ 30 for $^{76}$Zn and $^{84}$Ge, respectively.
The upper panel depicts the situation at low stellar densities
ranging from 10 - 10$^{4}$ g/cm$^{3}$ (the $\beta$-decay rates
remain constant in this density range), the middle panel for
intermediate stellar density 10$^{7}$ g/cm$^{3}$ and the lower panel
at a high stellar density of 10$^{11}$ g/cm$^{3}$. It is to be noted
that for both figures the abscissa is given in logarithmic scales.
Fig.~\ref{figure2} depicts stellar $\beta$-decay rates for $^{76}$Zn
in units of $s^{-1}$. It is to be noted that contribution from all
excited states are included in the final calculation of all decay
rates (refer to Eq.~21). It can be seen from this figure that, for
low and intermediate densities, the allowed $\beta$-decay rates are
one order of magnitude bigger at low temperatures and around two
orders of magnitude bigger at T$_{9}$ = 30. The phase space factor
of allowed GT is around one order of magnitude bigger than U1F phase
space for $^{76}$Zn (see Fig.~\ref{figure4}). At high densities the
allowed $\beta$-decay rates are several orders of magnitude bigger
at low temperatures and around two orders of magnitude bigger at
T$_{9}$ = 30. The reason primarily lies in the calculation of phase
spaces at high densities which we discuss shortly. The U1F
$\beta$-decay rates are around a factor six smaller at low
temperatures and densities for $^{84}$Ge (Fig.~\ref{figure3}). At
high temperatures allowed rates are a factor 50 bigger. As density
increases the allowed $\beta$-decay rates become several orders of
magnitude bigger at low temperatures and more than one order of
magnitude bigger at T$_{9}$ = 30.

We present our calculated stellar $\beta$-decay rates (GT and U1F
contributions) for remaining Zn and Ge isotopes in Tab.~\ref{ta1}
and Tab.~\ref{ta2}, respectively. In these tables entries $1.00
\times 10^{-100}$ represent decay rates smaller than $1.00 \times
10^{-100}$ s$^{-1}$.

\begin{figure}[htbp]
\begin{center}
\includegraphics[width=1.0\textwidth]{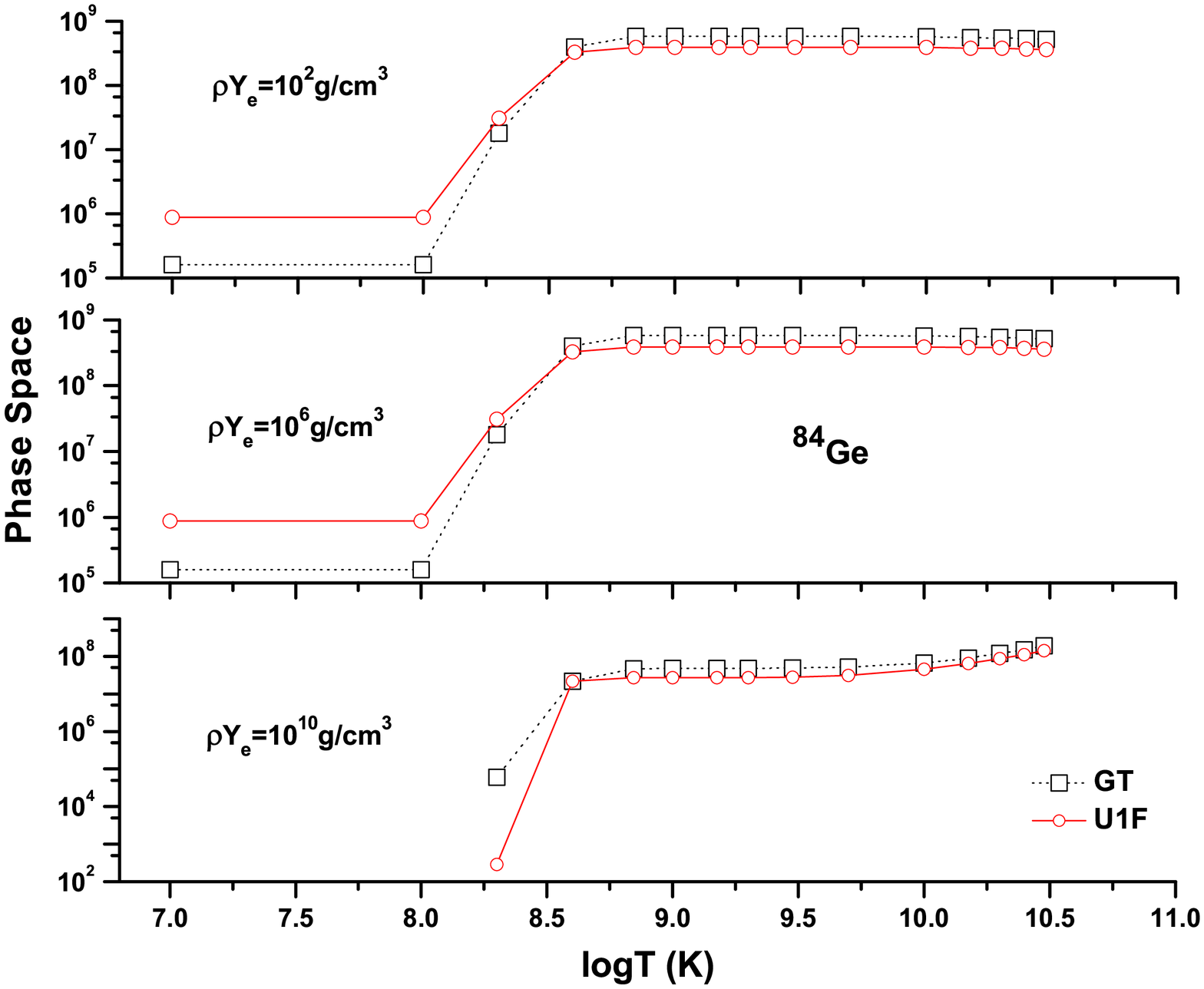}
\caption{\scriptsize Same as Fig.~\ref{figure4} but for
$^{84}$Ge.}\label{figure5}
\end{center}
\end{figure}

The amplification of phase space under stellar conditions due to U1F
transitions results in significant enhancement in the calculated
total $\beta$-decay rates. The phase space integrals for U1F
transitions compete well with those of allowed GT and under certain
stellar conditions supersede the allowed phase space (at low stellar
temperatures). The phase space calculation for allowed and U1F
transitions, as a function of stellar temperature and density, for
the neutron-rich $^{76}$Zn and $^{84}$Ge is shown in
Figs.~\ref{figure4} and ~\ref{figure5}, respectively. The phase
space is calculated at selected density of 10$^{2}$ g/cm$^{3}$,
10$^{6}$ g/cm$^{3}$ and 10$^{10}$ g/cm$^{3}$ (corresponding to low,
intermediate and high stellar densities, respectively) and stellar
temperature range T$_{9}$ = 0.01 - 30.  The phase space factors for
GT transitions start increasing at a much faster rate than those of
U1F transitions as stellar temperature increases. At T$_{9}$ = 30,
the GT phase space is much bigger. At high stellar densities the
phase space gets choked and becomes finite only as stellar
temperature soars to T$_{9}$ = 0.2. Similar trend in phase space
calculation is seen for heavier isotopes of Zn and Ge. The phase
space factor for U1F transitions gets amplified with increasing
neutron number for a particular element. These enlarged U1F phase
space contributes effectively to the forbidden $\beta$-decay rates
which we discussed earlier.

\begin{figure}[htbp]
\begin{center}
\includegraphics[width=1.0\textwidth]{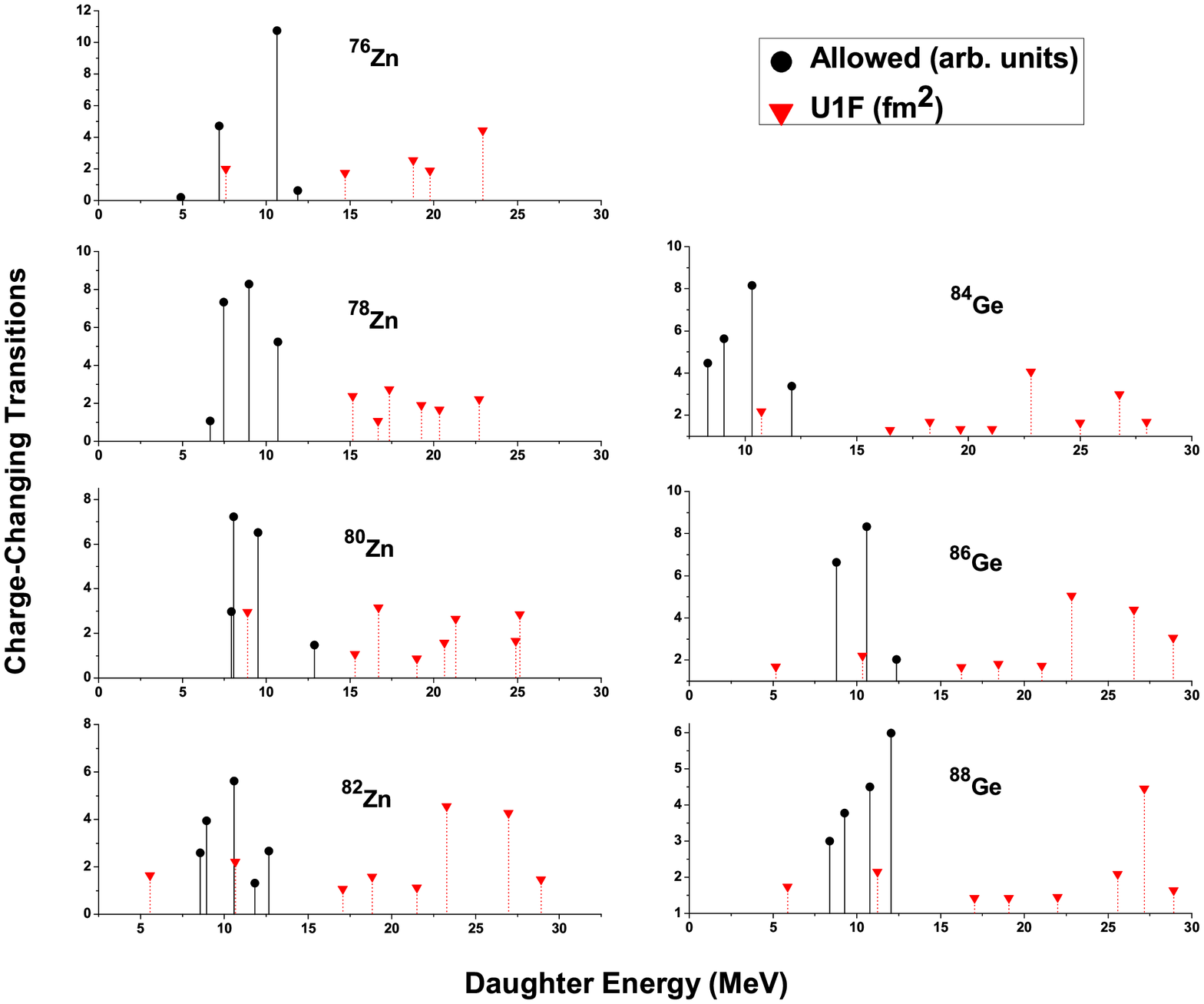}
\caption{\scriptsize Calculated strength distributions for allowed
GT ($0^{+} \rightarrow 1^{+}$) and unique FF transitions ($0^{+}
\rightarrow 2^{-}$) for Zn and Ge isotopes using the pn-QRPA(WS)
model. }\label{figure6}
\end{center}
\end{figure}

Towards the end of this section we present the calculated
charge-changing transition distributions. The calculated GT and U1F
transitions using the pn-QRPA(WS) model is shown in
Fig.~\ref{figure6} for Zn and Ge isotopes. As mentioned earlier all
these charge-changing transitions were quenched by a factor of 0.6.
It can be seen form this figure that the pn-QRA(WS) model calculates
high-lying transitions, specially for the U1F case. The transitions
shown at such high energies do not show much fragmentation which is
a drawback of neglecting deformation of nuclei.

\begin{figure}[htbp]
\begin{center}
\includegraphics[width=1.0\textwidth]{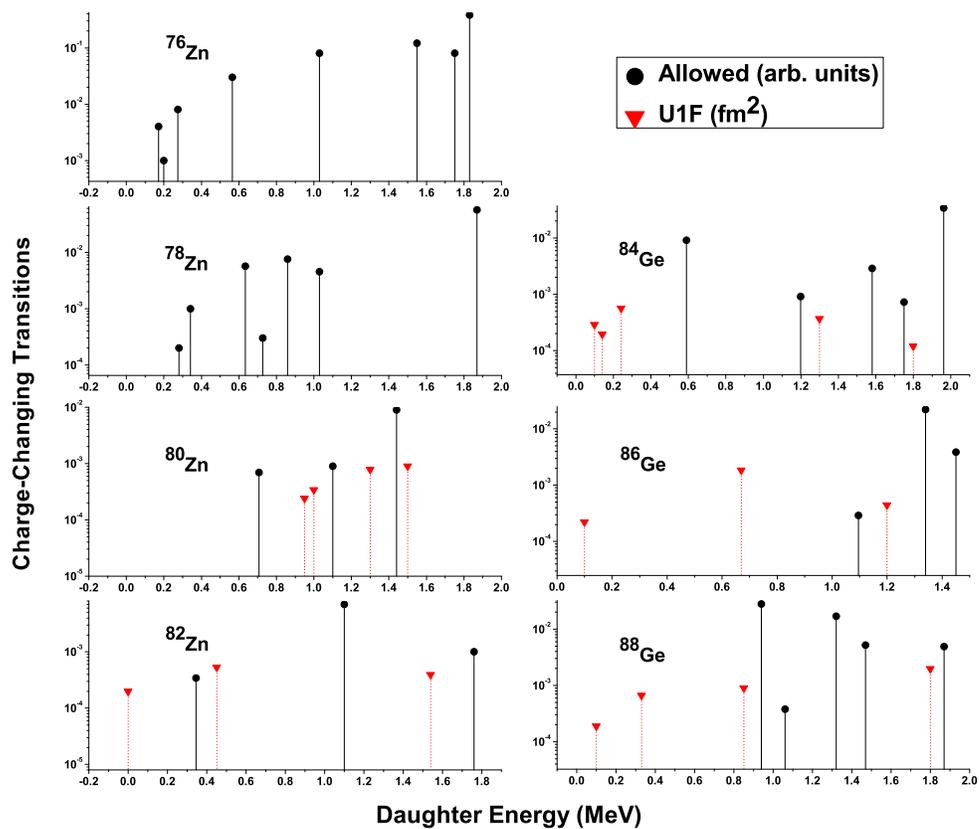}
\caption{\scriptsize Same as Fig.~\ref{figure6} but using the
pn-QRPA(N) model. }\label{figure7}
\end{center}
\end{figure}

It is a well known fact that taking  the deformation of the nucleus
into consideration results in the fragmentation of $\beta$-decay
strength distribution in the pn-QRPA formalism \cite{Hir93,Sta90}.
This also leads to shifting considerable strength in low-lying
daughter states. This is a big advantage of the pn-QRPA(N) model
that it is able to study the effects of deformation in calculation
of strength functions. To improve the reliability of calculated
transitions in our model, Experimental Unevaluated Nuclear Data List
(XUNDL) were incorporated in our calculation wherever possible.
Calculated excitation energies were replaced with measured levels
when they were within 0.5 MeV of each other. Missing measured states
were inserted and inverse transitions (along with their log$ft$
values), where applicable, were also taken into account. No
theoretical levels were replaced with the experimental ones beyond
the excitation energy for which experimental compilations had no
definite spin and/or parity assignment. This recipe for
incorporation of XUNDL data is same as used in earlier pn-QRPA
calculations of weak rates \cite{Nab99,Nab04}. The results of
calculated allowed and U1F transition strengths using the pn-QRPA(N)
model are shown in Fig.~\ref{figure7} for Zn and Ge isotopes. In
this figure the abscissa indicates the excitation energy in daughter
nuclei in units of MeV and a quenching factor of 0.6 is taken into
account for all charge-changing transitions. In Fig.~\ref{figure7}
we only show low-lying strength distribution up to excitation energy
of 2 MeV in daughter nucleus. It is to be noted that calculation was
performed up to 15 MeV in daughter. Calculated values of
charge-changing strength smaller than 10$^{-5}$ are not shown in
Fig.~\ref{figure7}. In contrast with the result of pn-QRPA(WS)
model, we notice low-lying allowed GT strength in daughter Ga and As
isotopes for reason mentioned above. The GT strength is also well
fragmented in the case of $^{76,78}$Zn. Whereas no U1F transition
strengths were calculated in $^{76,78}$Zn up to 2 MeV, we do
calculate these in the case of $^{80,82}$Zn. We do find a U1F
transition to ground state of $^{82}$Ga albeit of a very small
strength.  We obtain similar results for Ge isotopes using the
pn-QRPA(N) model. One should note the low-lying U1F transitions in
daughter arsenic isotopes. The U1F transitions result in reducing
the calculated $\beta$-decay half-lives bringing them in better
comparison with the experimental data (discussed earlier).

\begin{figure}[htbp]
\begin{center}
\includegraphics[width=1.0\textwidth]{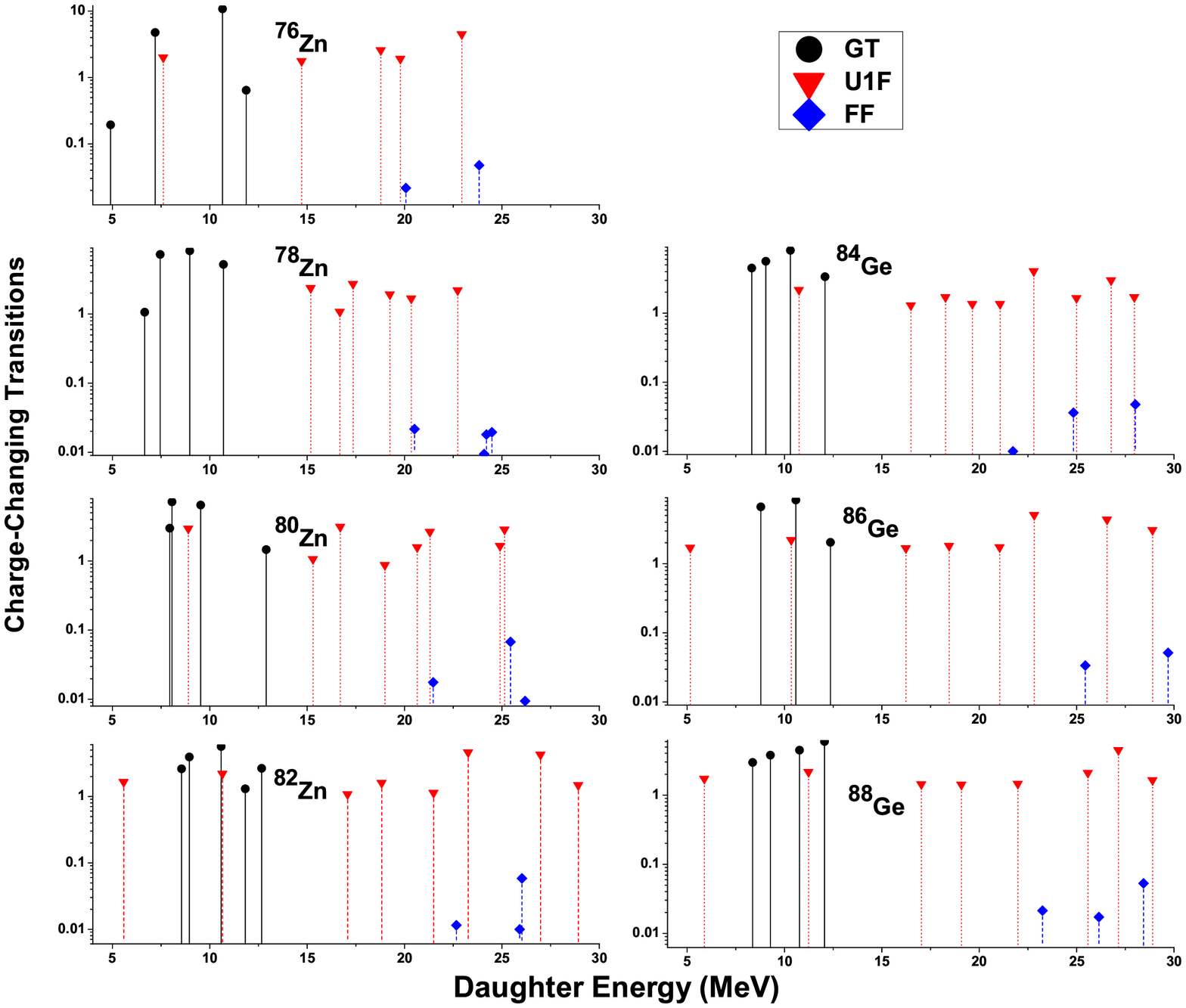}
\caption{\scriptsize Calculated strength distributions for allowed
GT ($0^{+} \rightarrow 1^{+}$), FF transitions ($0^{+} \rightarrow
0^{-}$) and unique FF transitions ($0^{+} \rightarrow 2^{-}$) for Zn
and Ge isotopes using the pn-QRPA(WS) model.}\label{figure8}
\end{center}
\end{figure}

Fig.~\ref{figure8} shows the calculation of GT, FF and U1F
transitions using the pn-QRPA(WS) model. As mentioned earlier we are
in a process of calculating FF transitions using the pn-QRPA(N)
model and this would be taken as a future assignment. It is seen
from Fig.~\ref{figure8} that FF transitions are placed at even
higher daughter energies. The results support the claim by
experiments performed by \cite{Hor80, Hor81} that the average energy
of the $0^{-}$ giant FF resonance lies at energy in excess of 20
MeV.

\section{Conclusions}
As the nuclei become heavier, strength of allowed GT transitions
gets smaller and with increasing neutron number the contribution of
FF transitions to the total half-lives becomes more significant. We
used two versions of the pn-QRPA model, one for spherical nuclei and
the other incorporating nuclear deformation, for our calculations.
The spherical pn-QRPA(WS) was used to calculate allowed GT (using
the Pyatov method), FF, U1F transitions and terrestrial
$\beta$-decay rates. The deformed pn-QRPA(N) was used to calculate
the allowed and U1F transitions as well as terrestrial and stellar
$\beta$-decay rates.

The half-life calculations were compared with the recent atomic mass
evaluation 2012 data and other theoretical calculations. The
inclusion of FF transitions improved the overall comparison of
calculated terrestrial $\beta$-decay half-lives in the pn-QRPA(WS)
model with the experimental data. Likewise, and more significantly,
the UIF contribution improved the comparison of pn-QRPA(N)
calculated half-lives. The DF3+CQRPA calculation was around a factor
two bigger than experimental data. The allowed GT calculation by
M\"{o}ller and collaborators was up to a factor 18 bigger but were
in  agreement with measured half-lives as neutron number increased
($^{82}$Zn and $^{88}$Ge).

Whereas the FF transitions to 0$^{-}$ and U1F transitions to 2$^{-}$
daughter states were calculated at rather high excitation energies
using the pn-QRPA(WS) model, the 2$^{-}$ states using the pn-QRPA(N)
were connected also to low-lying daughter states. Further the
spectra was more fragmented which was attributed to the deformation
parameter incorporated in the pn-QRPA(N) model.

It was shown that the U1F phase space has a sizeable contribution to
the total phase space at stellar temperatures and densities. It was
concluded that, for a particular element, the U1F phase space gets
amplified with increasing neutron number. Specially for the case of
$^{86,88}$Ge the calculated U1F phase spaces were several orders of
magnitude bigger and resulted in significant reduction of the
$\beta$-decay half-lives in comparison to what would result from
taking the contribution from the GT transitions alone into account.
For $^{86,88}$Ge roughly half the contribution to the total decay
rate comes from U1F transitions. This is a significant finding of
the current work. It is expected that contribution of FF transition
can increase further with increasing neutron number. The microscopic
calculation of U1F stellar $\beta$-decay rates, presented in this
work, could lead to a better understanding of the nuclear
composition and $Y_{e}$ in the core prior to collapse and collapse
phase.

New generation radioactive ion-beam facilities (e.g. FAIR (Germany),
FRIB (USA) and FRIB (Japan)) will make assessable measured strength
distribution of many more neutron-rich nuclei which in turn would
pose new challenges for theoretical models. Our results support the
argument that the pn-QRPA model calculates reliable half-lives for
exotic nuclei, specially near neutron drip-line. The reduced
$\beta$-decay half-lives bear consequences for the nucleosynthesis
problem and the site-independent $r$-process calculations. Our
findings might result in speeding-up of the $r$-matter flow relative
to calculations based on half-lives calculated  only from allowed GT
transitions. The effects of shorter half-lives resulted in shifting
of the third peak of the abundance of the elements in the
$r$-process toward higher mass region \cite{Suz11, Suz12}.
Simulators are urged to check for possible interesting changes in
their results by incorporating the reported rates. We are in a
process of including rank 1 operators in our FF calculation of
terrestrial $\beta$-decay half-lives in near future for still better
results. The allowed and U1F stellar $\beta$-decay rates on Zn and
Ge isotopes were calculated on a fine temperature-density grid,
suitable for simulation codes, and may be requested as ASCII files
from the corresponding author.

\ack N. Cakmak would like to thank C. Selam for very fruitful
discussion on calculation of $0^{+} \rightarrow 0^{-}$ transitions.
S. Stoica and J.-U. Nabi would like to acknowledge the support of
the Horia Hulubei Foundation and Romanian Ministry of National
Education, CNCS UEFISCDI, project PCE-2011-3-0318, Contract no.
58/28.10/2011. J.-U. Nabi would like to acknowledge the support of
the Higher Education Commission Pakistan through the HEC Project No.
20-3099.

\clearpage

\clearpage
\clearpage
\clearpage \clearpage
\clearpage

\clearpage


\begin{table}[h]
\caption{Allowed (GT) and  unique first-forbidden (U1F)
$\beta$-decay rates of $^{78,80,82}$Zn as a function of stellar
temperatures and densities. All stellar $\beta$ decay rates
($\lambda^{\beta}$) are given in units of sec$^{-1}$. Temperatures
(T$_{9}$) are given in units of 10$^{9}$ K. Densities
($\rho$Y$_{e}$) are given in units of g/cm$^{3}$.} \label{ta1} \tiny
\begin{tabular}{c|c|c|c|c|c|c|c|c|c}
Nucleus & T$_{9}$(K) & \multicolumn{4}{c|}{$\lambda^{\beta}$(s$^{-1}$)(Allowed)} & \multicolumn{4}{c}{$\lambda^{\beta}$(s$^{-1}$)(U1F)} \\
\cline{3-10} & & $\rho$$\it Y_{e}$=10$^{3}$ & $\rho$$\it
Y_{e}$=10$^{6}$ & $\rho$$\it Y_{e}$=10$^{9}$ & $\rho$$\it
Y_{e}$=10$^{11}$ & $\rho$$\it Y_{e}$=10$^{3}$ & $\rho$$\it
Y_{e}$=10$^{6}$ & $\rho$$\it Y_{e}$=10$^{9}$ & $\rho$$\it
Y_{e}$=10$^{11}$ \\

\hline
 & 0.01 & 3.22$\times10^{-1}$ & 3.11$\times10^{-1}$ &  1.00$\times10^{-100}$ & 1.00$\times10^{-100}$ & 7.79$\times10^{-2}$ & 7.74$\times10^{-2}$ & 1.00$\times10^{-100}$ & 1.00$\times10^{-100}$  \\
 & 1 & 3.22$\times10^{-1}$ & 3.12$\times10^{-1}$ & 8.61$\times10^{-7}$ & 3.72$\times10^{-91}$ & 7.79$\times10^{-2}$ & 7.74$\times10^{-2}$ & 6.92$\times10^{-7}$ & 1.00$\times10^{-100}$ \\
 $^{78}$Zn& 5 & 3.22$\times10^{-1}$ & 3.15$\times10^{-1}$ & 2.65$\times10^{-3}$ & 1.22$\times10^{-19}$ & 7.79$\times10^{-2}$ & 7.74$\times10^{-2}$ & 1.18$\times10^{-3}$ & 1.30$\times10^{-22}$ \\
 & 10 & 8.47$\times10^{-1}$ & 8.47$\times10^{-1}$ & 2.45$\times10^{-1}$ & 1.49$\times10^{-9}$ & 1.11$\times10^{-1}$ & 1.11$\times10^{-1}$ & 1.30$\times10^{-2}$ & 3.80$\times10^{-12}$ \\
 & 20 & 9.91$\times10^{0}$ & 9.91$\times10^{0}$ & 6.97$\times10^{0}$ & 4.48$\times10^{-4}$ & 3.39$\times10^{-1}$ & 3.39$\times10^{-1}$ & 1.66$\times10^{-1}$ & 2.57$\times10^{-6}$ \\
 & 30 & 2.01$\times10^{1}$ & 2.01$\times10^{1}$ & 1.73$\times10^{1}$ & 3.06$\times10^{-2}$ & 4.89$\times10^{-1}$ & 4.89$\times10^{-1}$ & 3.77$\times10^{-1}$ & 2.66$\times10^{-4}$ \\
\hline
 & 0.01 & 1.02$\times10^{0}$ & 9.98$\times10^{-1}$ & 3.35$\times10^{-4}$ & 1.00$\times10^{-100}$ & 1.47$\times10^{-1}$ & 1.47$\times10^{-1}$ & 1.41$\times10^{-4}$ & 1.00$\times10^{-100}$  \\
 & 1 & 1.02$\times10^{0}$ & 9.99$\times10^{-1}$ & 6.38$\times10^{-4}$ & 5.51$\times10^{-86}$ & 1.47$\times10^{-1}$ & 1.47$\times10^{-1}$ & 2.62$\times10^{-4}$ & 2.69$\times10^{-98}$ \\
 $^{80}$Zn& 5 & 1.02$\times10^{0}$ & 1.00$\times10^{0}$ & 1.33$\times10^{-2}$ & 1.86$\times10^{-18}$ & 1.47$\times10^{-1}$ & 1.47$\times10^{-1}$ & 3.95$\times10^{-3}$ & 5.22$\times10^{-22}$ \\
 & 10 & 1.76$\times10^{0}$ & 1.76$\times10^{0}$ & 4.58$\times10^{-1}$ & 6.02$\times10^{-9}$ & 1.78$\times10^{-1}$ & 1.78$\times10^{-1}$ & 2.32$\times10^{-2}$ & 7.46$\times10^{-12}$ \\
 & 20 & 1.92$\times10^{1}$ & 1.92$\times10^{1}$ & 1.36$\times10^{1}$ & 1.16$\times10^{-3}$ & 7.13$\times10^{-1}$ & 7.13$\times10^{-1}$ & 3.56$\times10^{-1}$ & 5.79$\times10^{-6}$ \\
 & 30 & 3.96$\times10^{1}$ & 3.96$\times10^{1}$ & 3.43$\times10^{1}$ & 6.99$\times10^{-2}$ & 1.17$\times10^{0}$ & 1.17$\times10^{0}$ & 9.08$\times10^{-1}$ & 6.65$\times10^{-4}$ \\
\hline
 & 0.01 & 4.58$\times10^{0}$ & 4.53$\times10^{0}$ & 1.67$\times10^{-1}$ & 1.00$\times10^{-100}$ & 1.21$\times10^{0}$ & 1.21$\times10^{0}$ & 1.26$\times10^{-1}$ & 1.00$\times10^{-100}$ \\
 & 1 & 4.58$\times10^{0}$ & 4.54$\times10^{0}$ & 1.72$\times10^{-1}$ & 5.96$\times10^{-70}$ & 1.21$\times10^{0}$ & 1.21$\times10^{0}$ & 1.27$\times10^{-1}$ & 1.67$\times10^{-84}$ \\
 $^{82}$Zn& 5 & 4.58$\times10^{0}$ & 4.54$\times10^{0}$ & 3.32$\times10^{-1}$ & 7.78$\times10^{-16}$ & 1.21$\times10^{0}$ & 1.21$\times10^{0}$ & 1.79$\times10^{-1}$ & 1.54$\times10^{-19}$ \\
 & 10 & 7.78$\times10^{0}$ & 7.76$\times10^{0}$ & 2.54$\times10^{0}$ & 9.82$\times10^{-8}$ & 1.53$\times10^{0}$ & 1.53$\times10^{0}$ & 4.19$\times10^{-1}$ & 2.70$\times10^{-10}$ \\
 & 20 & 5.77$\times10^{1}$ & 5.77$\times10^{1}$ & 4.17$\times10^{1}$ & 5.28$\times10^{-3}$ & 5.52$\times10^{0}$ & 5.52$\times10^{0}$ & 3.29$\times10^{0}$ & 7.76$\times10^{-5}$ \\
 & 30 & 1.07$\times10^{2}$ & 1.07$\times10^{2}$ & 9.29$\times10^{1}$ & 2.32$\times10^{-1}$ & 8.89$\times10^{0}$ & 8.89$\times10^{0}$ & 7.23$\times10^{0}$ & 6.83$\times10^{-3}$ \\
\hline
\end{tabular}
\end{table}

\begin{table}[h]
\caption{Same as Table~\ref{ta1} but for $^{86,88}$Ge.} \label{ta2}
\tiny \begin{tabular}{c|c|c|c|c|c|c|c|c|c}
Nucleus & T$_{9}$(K) & \multicolumn{4}{c|}{$\lambda^{\beta}$(s$^{-1}$)(Allowed)} & \multicolumn{4}{c}{$\lambda^{\beta}$(s$^{-1}$)(U1F)} \\
\cline{3-10} & & $\rho$$\it Y_{e}$=10$^{3}$ & $\rho$$\it
Y_{e}$=10$^{6}$ & $\rho$$\it Y_{e}$=10$^{9}$ & $\rho$$\it
Y_{e}$=10$^{11}$ & $\rho$$\it Y_{e}$=10$^{3}$ & $\rho$$\it
Y_{e}$=10$^{6}$ & $\rho$$\it Y_{e}$=10$^{9}$ & $\rho$$\it
Y_{e}$=10$^{11}$ \\

\hline
 & 0.01 & 1.62$\times10^{0}$ & 1.60$\times10^{0}$ & 2.21$\times10^{-1}$ & 1.00$\times10^{-100}$ & 6.84$\times10^{-1}$ & 6.84$\times10^{-1}$ & 1.92$\times10^{-1}$ & 1.00$\times10^{-100}$ \\
 & 1 & 1.62$\times10^{0}$ & 1.60$\times10^{0}$ & 2.22$\times10^{-1}$ & 1.02$\times10^{-76}$ & 6.84$\times10^{-1}$ & 6.84$\times10^{-1}$ & 1.93$\times10^{-1}$ & 1.23$\times10^{-82}$ \\
 $^{86}$Ge& 5 & 1.63$\times10^{0}$ & 1.63$\times10^{0}$ & 2.85$\times10^{-1}$ & 5.75$\times10^{-17}$ & 6.87$\times10^{-1}$ & 6.87$\times10^{-1}$ & 2.19$\times10^{-1}$ & 1.05$\times10^{-18}$ \\
 & 10 & 4.16$\times10^{0}$ & 4.16$\times10^{0}$ & 1.79$\times10^{0}$ & 2.87$\times10^{-8}$ & 1.07$\times10^{0}$ & 1.07$\times10^{0}$ & 4.57$\times10^{-1}$ & 6.76$\times10^{-10}$ \\
 & 20 & 2.49$\times10^{1}$ & 2.49$\times10^{1}$ & 1.79$\times10^{1}$ & 1.61$\times10^{-3}$ & 3.92$\times10^{0}$ & 3.92$\times10^{0}$ & 2.65$\times10^{0}$ & 9.51$\times10^{-5}$ \\
 & 30 & 3.94$\times10^{1}$ & 3.94$\times10^{1}$ & 3.41$\times10^{1}$ & 7.06$\times10^{-2}$ & 5.66$\times10^{0}$ & 5.66$\times10^{0}$ & 4.79$\times10^{0}$ & 5.96$\times10^{-3}$ \\
\hline
 & 0.01 & 4.63$\times10^{0}$ & 4.61$\times10^{0}$ & 1.63$\times10^{0}$ & 1.00$\times10^{-100}$ & 3.40$\times10^{0}$ & 3.39$\times10^{0}$ & 1.96$\times10^{0}$ & 1.00$\times10^{-100}$ \\
 & 1 & 4.63$\times10^{0}$ & 4.62$\times10^{0}$ & 1.64$\times10^{0}$ & 3.29$\times10^{-67}$ & 3.40$\times10^{0}$ & 3.39$\times10^{0}$ & 1.96$\times10^{0}$ & 2.42$\times10^{-71}$ \\
$^{88}$Ge& 5 & 4.68$\times10^{0}$ & 4.68$\times10^{0}$ & 1.79$\times10^{0}$ & 4.15$\times10^{-15}$ & 3.42$\times10^{0}$ & 3.41$\times10^{0}$ & 2.03$\times10^{0}$ & 5.01$\times10^{-16}$ \\
 & 10 & 9.25$\times10^{0}$ & 9.25$\times10^{0}$ & 4.78$\times10^{0}$ & 1.86$\times10^{-7}$ & 5.35$\times10^{0}$ & 5.35$\times10^{0}$ & 3.44$\times10^{0}$ & 2.75$\times10^{-8}$ \\
 & 20 & 3.89$\times10^{1}$ & 3.89$\times10^{1}$ & 2.88$\times10^{1}$ & 3.45$\times10^{-3}$ & 1.84$\times10^{1}$ & 1.84$\times10^{1}$ & 1.44$\times10^{1}$ & 1.12$\times10^{-3}$ \\
 & 30 & 5.74$\times10^{1}$ & 5.74$\times10^{1}$ & 5.01$\times10^{1}$ & 1.19$\times10^{-1}$ & 2.69$\times10^{1}$ & 2.69$\times10^{1}$ & 2.39$\times10^{1}$ & 4.91$\times10^{-2}$ \\
\hline
\end{tabular}
\end{table}

\end{document}